\documentstyle[aps,twocolumn]{revtex}
\newcommand{\be}{\begin{equation}}
\newcommand{\ee}{\end{equation}}
\newcommand{\bear}{\begin{eqnarray}}
\newcommand{\eear}{\end{eqnarray}}

\def\la{\mathrel{\mathpalette\fun <}}
\def\ga{\mathrel{\mathpalette\fun >}}
\def\fun#1#2{\lower3.6pt\vbox{\baselineskip0pt\lineskip.9pt
 \ialign{$\mathsurround=0pt#1\hfil##\hfil$\crcr#2\crcr\sim\crcr}}}

\begin{document}
\draft
\wideabs{
\title{  Determination of $\alpha_s(1 \;GeV)$ from the charmonium fine
structure.}
\author{\mbox{A.M.~Badalian\footnotemark{}, 
V.L.~Morgunov}\footnotetext{E-mail:badalian@vxitep.itep.ru}}
\address{Institute for Theoretical and Experimental Physics, \\
  RU-117218,  Moscow, B.Cheremushkinskaya 25, Russia}
\date{\today}
\maketitle

\begin{abstract}

The strong coupling constant $\alpha_s(\mu)$ is extracted from
the fits to charmonium spectrum and fine structure splittings.
The relativistic kinematics is taken into account and
relativistic corrections are shown to increase the matrix
elements defining spin effects up to $40\%$. The value of
$\alpha_s(\mu)$ at low--energy scale $\mu=1.0 \pm 0.2 \;GeV$ was
found to be $\alpha_s(\mu) = 0.38 \pm 0.03 (exp.) + 0.04
(theor.)$ which is about $50\%$ lower than standard perturbative
two--loop approximation  and is in good agreement with the
freezing $\alpha_s$ behaviour.

\end{abstract}
}
\section{Introduction}

The strong coupling constant $\alpha_s(Q^2)$ was measured in many
experiments at large momenta where perturbative QCD can be
successfully applied \cite{1}. Due to these investigations our
knowledge of the fundamental constant $\Lambda$ in QCD at high
energy scale is now much more precise than before. At the low
energy scale any extraction of $\alpha_s$ from experiment is
complicated by essential nonperturbative (NP) effects and the
$\tau$ decay is assumed to be the only lowest--energy process
from which the coupling constant $\alpha_s$ can be cleanly
extracted: for $\tau$ decay NP are argued to effects give a
small contribution to the hadronic ratio $R_{\tau}$ \cite{2}.
The averaged value of $\alpha_s$ at the scale of $\tau$ lepton 
mass is now $\alpha_s(M_{\tau}) = 0.35 \pm 0.03$ \cite{1}.

In the present paper we suggest another piece of low--energy
data -- very precise experimental measurements of spin splittings
of $\chi_c$ mesons. For reasons discussed below these data can
give a unique information on the strong coupling constant
$\alpha_s(\mu)$ at low energy scale, $\mu \la 1 \;GeV$. Even
though NP effects are essential for $1P$ charmonium state with
the size as large as $R \approx 0.65 \;fm$ the NP contributions
will be shown to affect $\alpha_s(\mu)$ in a simple and
controlled way and the choice of NP parameters is strongly
restricted by a fit to charmonium spectrum.

We shall not consider here the hyperfine shift of $\chi_c$ meson
with regard to the center of gravity of $~^3P_j$ multiplet. This
shift is small and negative, $|\Delta_{HF}| < 1 \;MeV$, and
small perturbative (P) and NP contributions cancel in $\Delta_{HF}$
\cite{3,4}. To explain this shift a very precise theory of NP
interaction at small distances is needed which is absent by now.

The experimental magnitude of tensor and spin--orbit splittings 
on the contrary are large enough $( > 35 \;MeV)$ and known with the
accuracy better than $2 \%$.

The spin structure of $P$ states in heavy quarkonia was already
investigated several times in the framework of potential approach
\cite{4,5,6,7}. We quote here the values of $\alpha_s(\mu)$
obtained in \cite{5,6} from the fit to charmonium fine structure:
$\alpha_s(1.22 \;GeV) = 0.386 (\mu \neq m, m = 1.30 \;GeV)$ in
\cite{5} and $\alpha_s(\mu = m = 1.20 \;GeV) = 0.35$ in
\cite{6}. Note that however different static interactions were
used in \cite{5,6}, nevertheless the resulting $\alpha_s(\mu)$
values at the scale $\mu \approx 1.2\;GeV$ are rather close to
each other and small in both cases. The calculated value of
$\alpha_s(1.2 \;GeV)$ is close to that of $\alpha_s(M_{\tau})$
at the energy scale that is significantly smaller than
$M_{\tau}$.

In this paper an improved analysis of the fine structure data
will be presented. First, we take into account relativistic
kinematics. The wave functions and all matrix elements will be
calculated with the help of spinless Salpeter equation as
compared to the nonrelativistic unperturbed Hamiltonian used in
\cite{5,6}. For the charmonium $1P$ state $v^2/c^2 \approx 0.4$
is not small and relativistic corrections are important. For
example, whereas the size of $1P$ state decreases only by $\sim
8\%$ in relativistic case, the matrix elements like $\langle
r^{-3} \rangle $, $\langle r^{-3}\ln mr\rangle $, defining spin
splittings, increase as much as $40\%$ for any set of
parameters. As a result relativistic corrections affect the
value of $\alpha_s(\mu)$ extracted from experimental data.

The second improvement is based on a simple observation. We
suggest instead of spin--orbit matrix element $a = \langle
V_{SO}(r)\rangle$ and tensor matrix element $c = \langle
V_{T}(r)\rangle $ to use their linear combination $\eta = 3/2 \;
c - a$. This combination has two remarkable properties:

$i)$ its perturbative part $\eta_P$ contains only $\alpha_s^2$
radiative corrections since $O(\alpha_s)$ terms cancel due to the
relation $c_{P}^{(1)} = 2/3 a_{P}^{(1)}$;

$ii)$ the parameter $\eta_P$ does not explicitly depend on the
renormalization scale mass $\mu$, whereas each of the second
order terms,  $c_{P}^{(2)}$ and $a_{P}^{(2)}$, contains a term
proportional to $\ln \mu / m$ (see Sect. 2).

Due to these features using $\eta$ instead of spin--orbit
matrix element $a$ is much more convenient while fitting to fine
structure data. The experimental values of matrix elements $a,c$
and $\eta$ can be easily calculated from $\chi_c$ meson masses
\cite{1}:
\bear 
   a_{exp} = 34.56 \pm 0.19 \;MeV, \nonumber \\
   c_{exp} = 39.12 \pm 0.62 \;MeV, \nonumber \\
   \eta_{exp} = 24.12 \pm 1.12 \;MeV
\label{eq1}
\eear
Note here that $\eta_{exp}$ is relatively large $(\approx 24
\;MeV)$ with a small experimental error $< 5\%$. The cited
numbers, Eq.~\ref{eq1}, slightly differ from the ones used in
\cite{6} because of recent change in $\chi_{c0}$ mass \cite{1}.

Our fitting procedure includes also a fit to charmonium spectrum.
Here we prefer more refined fitting to mass level differences than
to the absolute values $M(nL)$ for the given state. The most
important mass differences of levels lying below the open charm 
threshold are
\bear  
   M(2S) - M(1S) = 595.39 \pm 1.91 \;MeV, \nonumber \\
   M(1P) - M(1S) = 457.92 \pm 1.0 \;MeV,  \nonumber \\
   M(2S) - M(1P) = 137.47 \pm 1.77 \;MeV
\label{eq2}
\eear
Here $M(nL)$ is a spin--averaged mass.

\section{Perturbative fine structure parameters}

Every matrix element $a,c$ or $\eta$ includes both P and NP
contributions: $c = c_P + c_{NP}$, etc. First, P interaction will
be discussed. The spin--dependent P interaction is now known in
one--loop approximation only. Therefore our analysis can be only
done with $O(\alpha_s^2)$ terms. In coordinate space the
spin--dependent interaction including $\alpha_s^2$ corrections in
the renormalization $\overline{MS}$ scheme was calculated in
\cite{8}. Matrix elements of spin--orbit and tensor interactions
for a number of flavors $n_f = 3$, can be easily found from the
potentials \cite{8},
\be  
   a_P = a_P^{(1)} + a_P^{(2)}, \;
   a_P^{(1)} = \frac{2 \alpha_s(\mu)}{m^2}\langle r^{-3}\rangle,
\label{eq3}
\ee
\bear 
  a_P^{(2)} = \frac{2 \alpha_s^2(\mu)}{\pi m^2}
  \left\{ 4.5\langle r^{-3}\rangle \ln\frac{\mu}{m} + \right. \nonumber \\
  \left. 1.58193\langle r^{-3}\rangle  + 2.5 \langle r^{-3}\ln(m r)\rangle,
  \right\}
\label{eq4}
\eear
\be 
   c_P = c_P^{(1)} + c_P^{(2)}, \;
   c_P^{(1)} = \frac{4}{3} \frac{\alpha_s(\mu)}{m^2}\langle r^{-3}\rangle.
\label{eq5}
\ee
\bear 
  c_P^{(2)} = \frac{4}{3} \frac{\alpha_s^2(\mu)}{\pi m^2}
  \left\{ 4.5\langle r^{-3}\rangle \ln\frac{\mu}{m} + \right. \nonumber \\
  \left. 3.44916\langle r^{-3}\rangle  + 1.5 \langle r^{-3}\ln(m r)\rangle
  \right\}
\label{eq6}
\eear
In Eqs.~(\ref{eq3}--\ref{eq6}) all matrix elements will be
calculated for $1P$ state. From Eq.~(\ref{eq4}) and
Eq.~(\ref{eq6}) one can see that $a_P^{(2)}$ and $c_P^{(2)}$
contain $\ln{\mu/m}$ with the large coefficient whenever $\mu
\ne m$. However in the linear combination $\eta_P = 3/2 c_P -
a_P$ these terms are cancelled and the following simple
expression is obtained for $\eta_P$,
\be 
   \eta_P = \frac{3}{2} c_P^{(2)} - a_P^{(2)} =
   \frac{2 f_0}{\pi m}\alpha_s^2(\mu),
\label{eq7}
\ee

where $f_0$ is the combination of matrix elements,
\be 
   f_0 = (m)^{-1}
   \left\{1.86723 \langle r^{-3}\rangle  - \langle r^{-3} \ln m r\rangle
   \right\},
\label{eq8}
\ee
which has a weak dependence on parameters of static interaction
and charm--quark mass. Practically in all cases $f_0 = 0.12 \pm
0.01$.

The simple connection between $\alpha_s^2(\mu)$ and $\eta_P$,
Eq.~(\ref{eq7}), will be used later in our fit to the experimental
fine structure parameter $\eta_{exp}$, given by Eq.~(\ref{eq1}).

\section{Static interaction}

All spin effects in charmonium are very small as compared to the
level masses or mass level differences in Eq.~(\ref{eq2}) and
therefore can be considered as a perturbation. At this point the
choice of unperturbed Hamiltonian is of importance. In \cite{6}
Schr\"odinger Eq. with the Cornell potential was used for this
purpose. Here instead we take relativistic spinless Salpeter
equation (SSE),
\be 
   \left[ 2 \sqrt{\hat p^2 +m^2} + V_0(r)\right] \psi_{nl}(r)
   = M_{nl} \psi_{nl}(r)
\label{eq9}
\ee
In the framework of potential model this equation with square
root kinetic term was successfully used in many calculations of
meson masses and properties during last twenty years \cite{9,10}.
Recently it was deduced directly from QCD under a assumption of
area law of Wilson loop in the framework of proper--time
Feynman--Schwinger approach \cite{11}. Therefore SSE,
Eq.~(\ref{eq9}), cannot be consider as an {\it ad hoc} potential
model, but rather on  the same grounds as the QCD sum rules
approach.

In Eq.~(\ref{eq9}) the charm--quark mass $m$ is a pole mass
defined by the pole of quark propagator \cite{7}. The pole mass
$m$ is related to the running mass in $\overline{MS}$ scheme, $m
(\overline{m}^2)$, as \cite{7}
\bear 
   m \equiv m_{pole} = \overline{m} (\overline{m}^2) \left\{ 1 +
   \frac{4}{3} \frac{\alpha_s(m^2)}{\pi} + \right. \nonumber \\
   \left.(K-\frac{8}{3})\left(\frac{\alpha_s(m^2)}{\pi}\right)^2  \right\}
\label{eq10}
\eear
For $c$ quark $K \approx 14.0$. From Eq.~(\ref{eq10}) one can
estimate that for example  for $m = 1.4 \;GeV$ and $\alpha_s(m)
\approx 0.3$ the relation $m/\overline{m} \approx 1.23$, i.e.
$m$ is about $20 \div 30 $ \% larger than $\overline{MS}$ mass
$\overline{m}$. The static potential $V_0(r)$ in Eq.~(\ref{eq9})
is taken here as Coulomb potential plus linear confining term,
as it was done in \cite{6},
\be 
   V_0(r) = -\frac{4}{3} \frac{\tilde{\alpha}_V(\mu)}{r} + \sigma r
\label{eq11}
\ee
Here the vector coupling constant  $\tilde{\alpha}_V(\mu)$
differs from the running constant $\alpha_s(\mu)$ in $\overline{MS}$
scheme. The connection between them was found in \cite{12} and
for $n_f = 3$ it reads
\be 
   \tilde{\alpha}_V(\mu) = \alpha_s(\mu) [ 1 +
   \frac{1.75}{\pi}\alpha_s(\mu) ]
\label{eq12}
\ee
In Eq.~(\ref{eq11}) $\tilde{\alpha}_V(\mu)$ will be taken at
some fixed point $\mu$ which in general can differ from the
scale $\mu_0$ which defines spin splittings because Coulomb
interaction behaves as $r^{-1}$ whereas spin--dependent
interaction behaves as $r^{-3}$ and therefore is more sensitive
to smaller distances. However, in our calculations it was found
that the choice $\mu \approx \mu_0$ together with the additional
condition, Eq.~(\ref{eq12}), gives rise to a good description of
both charmonium spectrum and fine structure.

In Eq.~(\ref{eq11}) confining potential was chosen to be linear 
at all distances, as it is done in most papers. This is in 
agreement with direct lattice calculations of static interaction 
\cite{14}, but is at variance with OPE and field correlator 
approach, which require that the NP interaction should start at 
small $r$ as $const \cdot r^2$ and tend to linear form $\sigma r$
only at $r > T_g$, where $T_g$ is the gluonic correlation length,
measured on the lattice in \cite{15} to be $T_g \approx 0.3 \;fm$.

Recently some arguments have been given in favour of additional
linear potential at small $r$ \cite{16} and it was explicitly
found from P--NP interference in \cite{17} to be of the
magnitude which effectively resolves the discrepancy and confirm
the linear NP potential at all $r$.

Therefore we assume below that linear potential $\sigma r$ is
valid at all distances. To control the dependence of our results
on the choice of string tension the parameter $\sigma$ will be
varied in some interval compatible with a good description of
mass level differences in charmonium.

For pure linear potential NP contribution to spin--orbit
potential is given by Thomas interaction for which
\be 
   a_{NP} = -\frac{\sigma}{2m^2} \langle r^{-1}\rangle , \; a_{NP} < 0
\label{eq13}
\ee
The value of $c_{NP}$ is defined by ${\cal{D}}_{1}$--correlator
\cite{13,18} which is small in lattice calculations \cite{15},
yielding the estimate $0 < c_{NP} < 0.3 \;MeV$ \cite{18}. This
result is in a good agreement with direct lattice calculations
of NP tensor potential \cite{19}, where $V_T(r)$ was found at
the distances $0.2 \la r \la 0.6 \;fm$ to be $V_T(r) < 1 \;MeV$.
Thus value of $c_{NP}$ is much less than $|a_{NP}| \sim 15 \div
20 \;MeV$ and therefore can be neglected in $\eta_{NP}$. The
theoretical error due to neglected term in Eq.~\ref{eq14} is
small and will be taken into account in our analysis. Then
\be 
   \eta_{NP} = \frac{3}{2} c_{NP} - a_{NP} 
   \approx \frac{\sigma}{2 m^2}\langle r^{-1}\rangle
\label{eq14}
\ee

To calculate wave functions of SSE the expansion of $\psi_{nl}$
in a series over Coulomb--type functions, suggested in \cite{9},
was used. The numerical calculations for $1P$ state provided
high accuracy (better than $10^{-3}$) for matrix elements like
$\langle r^{-1}\rangle$, $\langle r^{-3}\rangle$, etc.. Some of
them are listed in Table~\ref{tab1} both for SSE and
nonrelativistic Schr\"odinger equation for three sets of
parameters. Set A with $m = 1.2 \;GeV$, $\sigma = 0.2 \;GeV^2$
and $\tilde{\alpha}_V = 0.35$ was taken from the paper \cite{6}.
From Table~\ref{tab1} one can see that the difference between
relativistic (R) and nonrelativistic (NR) matrix elements is
about 8\% for square root radius, $\sim 10\%$  for matrix
element $\langle r^{-1}\rangle$, defining NP effects,
Eq.~(\ref{eq13}), and very large, $\sim 40\%$, for matrix
element $\langle r^{-3}\rangle $. This growth of $\langle
r^{-3}\rangle $ is due to the decreasing $1P$--state size in
relativistic case.

One should also note here that mass difference $M(2S) - M(1P)$
is large in NR case $( \sim 170 \;MeV)$, i.e. $\sim 20\%$ larger
than the experimental value $137 \;MeV$, Eq.~(\ref{eq2}). When
relativistic kinematic is taken into account then excellent
agreement with experiment can be easily reached for this mass
difference, see Table~\ref{tab2}.

\section{Fitting conditions}

Our fitting procedure includes two conditions:
(I) $ \eta = \eta_{exp}$ and (II) $c = c_{exp}$. Using 
Eqs.~(\ref{eq7}) and (\ref{eq14}) we write the first condition as
\bear 
   \eta = \frac{3}{2}c_P^{(2)} - a_P^{(2)} +
   \frac{\sigma}{2m^2}\langle r^{-1}\rangle  = \nonumber \\
    = \eta_{exp} = (24.12 \pm 1.12) 10^{-3} \;GeV
\label{eq15}
\eear
With the help of Eqs.~(\ref{eq4}) and (\ref{eq6}) 
this condition can be rewritten in another form,
\FL
\be 
   \alpha_s^2(\mu) \cdot \frac{2 f_0}{\pi m} =
   (24.12\pm 1.12) 10^{-3} - \frac{\sigma}{2 m^2}\langle r^{-1}\rangle
   \equiv \Delta
\label{eq16}
\ee
From here $\alpha_s(\mu)$ can be defined through the known
numbers,
\be 
   \alpha_s(\mu) = \sqrt{\frac{\pi m \Delta}{2 f_0}}
\label{eq17}
\ee
where $m,\sigma$ are fixed parameters and $f_0$ is given by the
expression in Eq.~(\ref{eq8}). Our calculations show that $f_0$
is almost constant, $f_0 \approx 0.12 \pm 0.01$, independently
on the choice of other parameters of static interaction, $m,
\sigma$ and $\tilde{\alpha}_V$.

As seen from Eq.~(\ref{eq17}) the coupling constant is
proportional to $\sqrt{m}$ and its value also strongly depends
on $\sqrt{\Delta}$, where $\Delta$ in Eq.~(\ref{eq16}) is the
difference between $\eta_{exp}$ and the absolute value of NP
spin--orbit matrix element $|a_{NP}|$. As it follows from our
calculations this difference can become negative for some small
quark masses $(m \la 1.3 \;GeV)$ and large $\sigma \approx 0.2
\;GeV^2$ whereas the l.h.s. of the Eq.~(\ref{eq16}) is always
positive. Therefore the solutions with such small masses and
large $\sigma$ must be excluded from our fit to the experimental
data. For example, the Set A from \cite{6} considered in the
relativistic case yields the parameter $|a_{NP}| = 26.53 \;MeV$
which is larger than $\eta_{exp}$ in Eq.~(\ref{eq1}) and hence
inappropriate.

With the use of Eq.~(\ref{eq17}) $\alpha_s(\mu)$ can be
precisely determined for the given set of parameters $m$ and
$\sigma$. At this stage the scale $\mu$ still remains undefined
but $O(\alpha_s)$ terms, $c_P^{(1)}$ and $a_P^{(1)}$,  which do
not explicitly depend on $\mu$, can be calculated.

At the second step one can fit the second condition, $c =
c_{exp} = 39.12 \pm 0.62 \;MeV$, where $c = c_P^{(1)} +
c_P^{(2)} + c_{NP}$. As it was discussed above NP term $c_{NP}$
is small,$c_{NP} < 0.3 \;MeV$, and as compared to $c$ can be
neglected. With $c_P^{(1)}$ already calculated above,
Eq.~(\ref{eq5}), the second condition can be rewritten as
\be 
   c_P^{(2)} = c_{exp} -
   \frac{4}{3}\frac{\alpha_s(\mu)}{m^2}\langle r^{-3}\rangle
\label{eq18}
\ee
Using Eq.~(\ref{eq6}) it can be represented as
\bear 
   \frac{4}{3}\frac{\alpha_s(\mu)}{\pi m^2}    \left\{
   \langle r^{-3}\rangle [3.44916 + 4.5 \ln\frac{\mu}{m}] + \right.\nonumber \\
    \left. 1.5 \langle r^{-3}\ln(m r)\rangle \right\} = \nonumber \\
   (39.12 \pm 0.62) 10^{-3} -
   \frac{4}{3}\frac{\alpha_s(\mu)}{m^2}\langle r^{-3}\rangle
\label{eq19}
\eear
All matrix elements and $\alpha_s(\mu)$ in Eq.~(\ref{eq19}) are
already known for given $m$, $\sigma$ and $\tilde{\alpha}_V$,
therefore Eq.~(\ref{eq19}) fixes $\mu$ at some value $\mu_0$. It
was found in our numerous calculations that the value $\mu_0$
does not coincide with the $c$ quark mass $m$ in general case,
and only for very special choice of charm quark mass and static
potential parameters $\tilde{\alpha}_V$ and $\sigma$ the
condition $\mu = m$ can be satisfied.
That might be the reason why the choice $\mu = m$ apriori taken
in the paper \cite{6} has difficulties with simultaneous fitting
of spin--orbit and tensor splittings (for $1P$ state) with the
same $\alpha_s(\mu)$ if $c$ quark mass $m = 1.8 \;GeV$. In our
fit, without the choice $\mu = m$, when the Eq.~(\ref{eq16}) and
Eq.~(\ref{eq19}) are satisfied, we "automatically" get $a =
a_{exp}$.

Hence for fixed $m$ and $\sigma$ one finds $\alpha_s(\mu)$ and
$\mu = \mu_0$ satisfying experimental data. In the next Section
we shall check whether this choice satisfies also data on the
whole spectroscopy of charmonium.

\section{The choice of parameters}

We shall discuss here only those fits with given
$m$, $\sigma$ and $\tilde{\alpha}_V$ which yield good description
of spin--averaged spectrum in charmonium. As a result there
appear some restrictions on the magnitude of $\sigma$ and
$\tilde{\alpha}_V$, but the choice of pole mass $m$ remains
relatively arbitrary. Even if one takes $\overline{MS}$ mass,
$\overline{m} = 1.30 \pm 0.20 \;GeV$, as it is commonly accepted
\cite{1}, and makes use of the Eq.~(\ref{eq10}), then the values
of the pole mass $m$ can vary from $m_{min} \approx 1.2 \;GeV$ to
$m_{max} \approx 1.8 \;GeV$. Therefore it is important to impose
additional physical restrictions on $m$. From our fitting
procedure it is clear that for any mass $m$ the corresponding
$\alpha_s(\mu_0)$ and $\mu_0$ can be formally found. However the
dependence of $\alpha_s(\mu)$ on $\mu$ is different for
different $m$. Some restrictions on the value of pole quark mass
come from our fit to the fine structure data.

All considered solutions with
different $m$ can be separated  in three groups: 

$i)$ Quark pole mass is small, $m \la 1.3 \;GeV$. Then in
Eq.~(\ref{eq16}) the difference $\Delta \approx 2 \div 4 \;MeV$
is also small since NP spin--orbit matrix element $|a_{NP}|$,
proportional to $m^{-2}$, is large. The value of $\Delta$ remains
small even for the small $\sigma \approx 0.17 \;GeV$. Then
according to Eq.~(\ref{eq17}) $\alpha_s(\mu) \sim \sqrt{m\Delta}$
is small, $\alpha_s(\mu) \la 0.20$. As a consequence,
$O(\alpha_s)$ terms, $c_P^{(1)}$ and $a_P^{(1)}$, are not large
compared to the second order terms $c_P^{(2)}$ and $a_P^{(2)}$
and in some cases $c_P^{(2)} > c_P^{(1)}$ were obtained. For
small $m$ the calculations give large value of scale, $\mu \ga
3\;GeV$, and in many cases $\mu \ga 10\;GeV$, so that in all
cases $\ln \mu/m \ga 1$. Note that the pole mass $m \la 1.3
\;GeV$ corresponds to the $\overline{MS}$ mass $\overline{m} \la
1.15 \;GeV$.

Therefore for small $m$ and any $\sigma$ the fine structure
parameters $a$ and $c$ strongly depend on $\mu$, and
$\alpha_s^2$ corrections give large or even dominant
contribution. Those solutions will be excluded in our analysis as
unphysical.

$ii)$ The pole mass $m$ is large, $m \ga 1.6 \;GeV$, which
corresponds $\overline{MS}$ mass $\overline{m} \ga 1.3 \;GeV$.
Then in contrast to small $m$ case the value $\Delta \approx 10
\div 12 \;MeV$ is large since the matrix element $|a_{NP}|$ is
smaller for large $m$. Therefore $\alpha_s(\mu) \sim \sqrt{m
\Delta}$ grows large,  $\alpha_s(\mu) \ga 0.42$ in our
calculations. However, for large $m$ the value of $\mu$ was
found to be small, $\mu \la 0.7 \;GeV$, so that $|\ln \mu/m|
\sim 1$ is large again. As a consequence the negative term,
proportional to $\ln \mu/m$ in $c_P^{(2)}$, Eq.~(\ref{eq6}), and
$a_P^{(2)}$, Eq.~(\ref{eq4}), cancel positive contribution from
other two terms, and as a result $O(\alpha_s^2)$ terms are
numerically rather small. In some cases $a_P^{(2)}$ even becomes
negative.

Thus the second order terms have strong dependence on $\mu$ and
one gets relatively large value of $\alpha_s(\mu) \approx 0.4
\div 0.5$  at small mass scale $\mu \la 0.7 \;GeV$. Therefore in
this case one can expect large contribution from $\alpha_s^3$
corrections which are still unknown and hence any decisive
conclusions about those solutions  with large $m$ are now
impossible.

$iii)$ The pole quark mass is in the range $1.4 \div 1.56
\;GeV$. For those masses $\mu$--dependent term in $c_P^{(2)}$
and $a_P^{(2)}$ does not dominate and typically $|\ln \mu/m| \la
0.4$ or $0.6 \la \mu/m \la 1.0$. Just for such masses the best
fit to charmonium spectrum was obtained.

From two fits, to the spectrum and the fine structure
splittings, we have found out the following restrictions on the
choice of $m$ and $\sigma$,

\be 
   m = 1.48 \pm 0.08 \;GeV, \;\; \sigma = 0.178 \pm 0.008 \;GeV^2
\label{eq20}
\ee
For such $m$ and $\sigma$ the Coulomb constant $\tilde{\alpha}_V$
is "automatically" fixed by the fit to charmonium spectrum:
$\tilde{\alpha}_V = 0.42 \pm 0.04$. Note also that the pole mass
$m$ in the range given by Eq.~(\ref{eq20}) corresponds to the
$\overline{MS}$ mass $\overline{m} \approx 1.18 \pm 0.07 \;GeV$.

With the parameters from Eq.~(\ref{eq20}) the calculated values of
$\mu_0$ are in range, $\mu_0 \approx 1.0 \pm 0.2 \;GeV$.
At this scale, $\mu_0 \sim 1 \; GeV$, the extracted value of 
$\alpha_s(\mu_0)$ (with experimental and theoretical errors) is

\be 
   \alpha_s(\mu_0) = 0.38 \pm 0.03(exp) \pm 0.04(theory)
\label{eq21}
\ee
\be 
   0.75 \le \mu_0 \le  1.2 \;GeV
\label{eq22}
\ee

Our calculations for two different sets of parameters are represented 
in Table~\ref{tab2}. 

Set D with $m=1.4\;GeV,\;\sigma=0.183\;GeV^2$, and
$\tilde{\alpha}_V=0.39$ was selected because for this variant
$\mu_0 \approx m$ was obtained, i.e. the term with $\ln\mu/m$
does not contribute to $c_P^{(2)}$ and $a_P^{(2)}$. Also for Set
D a good description of charmonium spectrum (see
Table~\ref{tab2}) was obtained. From the fine structure fit the
extracted $\alpha_s(\mu_0)$ and $\mu_0$ values are
\be 
   \alpha_s(\mu_0) = 0.312,\;\mu_0 = 1.36\;GeV\approx m
\label{eq23}
\ee
This $\alpha_s(\mu_0)$ is small as compared with
$\alpha_s(M_{\tau})$ extracted from $\tau$ decay \cite{1,2}.
If we put the additional restriction:
\be 
   \alpha_s(m < M_{\tau}) > \alpha_s(M_{\tau}) = 0.35 \pm 0.03
\label{eq24}
\ee
then we have to conclude that for Set D the Eq.~(\ref{eq24}) is
violated.

To control the dependence of $\alpha_s(\mu)$ on NP parameter
$\sigma$ we have calculated $\alpha_s(\mu)$ for smaller
$\sigma$, varying $\sigma$ till the description of charmonium
spectrum was becoming poor. For  example , for $\sigma = 0.17
\;GeV^2$ $M(2S) - M(1S)$ and $M(1P) - M(1S)$ were already $30
\;MeV$ lower than its experimental values. For this $\sigma =
0.17 \;GeV^2$ ($m = 1.40\;GeV$ and  $\tilde{\alpha}_V=0.39$) the
extracted $\alpha_s(\mu_0)$ is increasing,
\be 
   \alpha_s(\mu_0) = 0.373,\;\mu_0 = 0.90\;GeV,
\label{eq25}
\ee
these fitted values lie exactly in the range given by
Eq.~(\ref{eq21}) and Eq.~(\ref{eq22}).

The best fit in our calculations was found for the set of parameters C,
\bear 
   Set \; C:\; m = 1.48 \;GeV, \nonumber \\
   \sigma = 0.18 \;GeV^2, \; \tilde{\alpha}_V=0.42.
\label{eq26}
\eear
Note that the pole mass $m = 1.48 \;GeV$ corresponds to
$\overline{m} \approx 1.17 \;GeV$. For Set C the excellent
agreement for the mass level differences in charmonium (see
Table~\ref{tab2}) was obtained and
\be 
   \alpha_s(\mu_0) = 0.365 \pm 0.027(exp),\;\mu_0 = 0.91\;GeV
\label{eq27}
\ee
To estimate the theoretical error coming from NP effects we have
varied $\sigma$ and analyzed the $\alpha_s$ dependence on
$\sigma$. For the smaller $\sigma$, $\sigma = 0.17 \;GeV^2$,
$\alpha_s(\mu)$ is increasing,
\be 
   \alpha_s(\mu_0) = 0.408 \pm 0.027(exp),\;\mu_0 = 0.94\;GeV
\label{eq28}
\ee
For this value of $\sigma$ the fit to spectrum is worse as
compared to $\sigma = 0.18 \;GeV^2$, in particular mass level
differences are about $5 \div 10\%$ less than the experimental
ones, Eq.~(\ref{eq2}). Therefore $\alpha_s(\mu)$ given by
Eq.~(\ref{eq28}) can be considered as the upper limit for the
coupling constant $\alpha_s(\mu)$ with $m = 1.48\;GeV$.
Averaging two numbers, Eq.~(\ref{eq27}) and Eq.~(\ref{eq28}),
yields the extracted value of $\alpha_s(\mu)$ in
Eq.~(\ref{eq21}).

One should be reminded here that the nonrelativistic analysis in
\cite{5} discussed in Introduction has given $\alpha_s(1.22) =
0.386$ (and larger $\mu=1.22 \;GeV$) close to our resulting
value, Eq.~(\ref{eq21}), however, in relativistic case the
fitted value of  $\alpha_s(\mu)$ would be smaller for the same
parameters $m$, $\sigma$ and $\tilde{\alpha}_V$ used in
nonrelativistic approach in \cite{5}. Here we would like also to
note a remarkable agreement between our result Eq.~(\ref{eq21})
and the value $\alpha_s(\mu_0) = 0.38 \pm 0.05  (\mu_0 = 0.93
\;GeV)$ extracted from the best overall fit to $2P$ state in
botomonium \cite{4,7}. This coincidence is probably not
occasional. Both systems, $2P$ $b \overline{b}$ state and $1P$ $c
\overline{c}$ state, have exactly the same size:
$\sqrt{\langle r^2 (b \overline{b})\rangle_{2P}} = 
\sqrt{\langle r^2 (c \overline{c})\rangle_{1P}} = 0.63 \pm 0.03 \;fm$.

The extracted value $\alpha_s(0.9) \approx 0.38$ turns out to be
smaller then the corresponding value in PQCD. If we take for
$n_f=5$, $\Lambda^{(5)} = 237^{+26}_{-24}\;MeV$ in two--loop
approximation from recent compilation \cite{1} and calculate
$\Lambda^{(4)}$ and $\Lambda^{(3)}$ from the matching conditions
at $\overline{MS}$ quark masses, $\overline{m}_b=4.3\;GeV$ and
$\overline{m}_c=1.3\;GeV$, then $\Lambda^{(4)} = 338^{+33}_{-31}
\;MeV$ and $\Lambda^{(3)} = 384.4^{+32}_{-30} \;MeV$. Then in
two--loop approximation $\alpha_s(1\;GeV) =
0.54^{+0.06}_{-0.05}$ is about a factor $1.5$ larger than
$\alpha_s(1\;GeV) \approx 0.36 \pm 0.03 \pm 0.03$ in our fit,
the behaviour of perturbative $\alpha_s(\mu)$ is shown as a band
in Fig.~1. Seven points from our fits with different $c$ quark
masses, $1.40 \le m \le 1.48$ are also shown in Fig.~1.

The variation of quark mass gives rise to changes in the values
of $\mu$ and as a result we get the dependence of  
$\alpha_s(\mu)$ on the scale $\mu$ in low--energy region. It is
important to underline that for $1P$ state in charmonium the
scale parameter $\mu$ lies in low--energy region, $\mu \la 1
\;GeV$, in all our fits (our seven points are shown in Fig.~1)
$\alpha_s(\mu) \la 0.42$ is in agreement with
Eq.~(\ref{eq21}). In all cases the extracted values of
$\alpha_s(\mu)$ are below those in perturbation theory.

How to interpret this result? We cannot accept the point of
view that the relatively small value of $\alpha_s(1\;GeV)$ found
here points out that $\Lambda^{(3)}$ should be smaller than the
cited above perturbative value of $\Lambda^{(3)} \approx 384
MeV$ which corresponds to $\alpha_s(M_Z) = 0.119 \pm 0.002$. In
our fit to the charmonium fine structure we have met a very
specific case when $\alpha_s(\mu)$  is extracted at low--mass
scale, $\mu \la 1\;GeV$. In this region the $\alpha_s(\mu)$
freezing phenomenon is expected \cite{20} which can affect the
resulting values of the strong coupling constant, as is discussed
below.

\section{ {$\alpha_s$} freezing}
Perturbative evolution of $\alpha_s$ at large distances is
physically modified due to NP fields, which create confining
strings (and hence NP mass parameter) and make an effective
large--distance cut--off in all loop integrals. This mechanism
leads to a specific regime of $\alpha_s$ evolution at small $Q$,
which is called freezing, and was studied both theoretically
\cite{20} and in experimental analysis \cite{21}.

Theoretically the freezing of the strong coupling constant can
be deduced if one takes into account the behaviour of running
coupling constant in  background vacuum fields \cite{20}.
Experimentally phenomenon of freezing was anticipated already for
some time (see review in \cite{21}).

The coupling constant, denoted as $\alpha_B(q^2)$, appears to be
an analytical function at all Euclidean momenta, because it
depends on $q^2+m_B^2$, where $m_B^2$ is a background mass. At
large $q^2 \gg \Lambda_B^2$ $\alpha_B(q^2)$ coincides with
perturbative $\alpha_s(q^2)$. It means that $\Lambda_B^{(5)} =
\Lambda^{(5)}$ and $\alpha_B(q^2)$ keeps the property of
asymptotic freedom. At small $q^2$, $\alpha_B(q^2)$ is freezing
at some constant value, which depends on the background mass
$m_B$ \cite{22}. The behaviour of $\alpha_B(q^2)$ in two--loop
approximation is given by the following approximate expression
\cite{20,22,23}:

\be 
   \alpha_B(q^2) = \frac{4\pi}{\beta_0 t_B}\left( 1 -
   \frac{\beta_1}{\beta_0^2}\frac{\ln t_B}{t_B} \right)
\label{eq29}
\ee
with $t_B = \ln{\frac{q^2+m_B^2}{\Lambda^2_B}}$

The background mass $m_B$ in Eq.~(\ref{eq29}) can depend on the
process considered and for static interquark interaction $m_B$
coincides with the lowest hybrid mass \cite{20,22}. Analytic and
lattice calculations \cite{24} predict the energy of the lowest
hybrid excitation in the interval $1 \div 1.5 \;GeV$. In that
follows we find the best fitting value $m_B = 1.1 \;GeV$, which
is well inside the predicted interval.

Since  $\Lambda_B^{(5)} = \Lambda^{(5)}$ and $m_B$ is fixed,
$\Lambda_B^{(4)}$ and $\Lambda_B^{(3)}$ can be calculated from
the matching conditions at the flavor thresholds
$(\overline{m}_b=4.3\;GeV$ and $\overline{m}_c=1.3\;GeV)$. The
obtained values of $\Lambda_B^{(4)}$ and $\Lambda_B^{(3)}$ turn
out to be very close to perturbative values. In two--loop
approximation $\Lambda_B^{(4)} = 339^{+33}_{-31} \;MeV$ is only
by $1 \;MeV$ larger than $\Lambda_B^{(4)}$ and $\Lambda_B^{(3)}
= 400^{+33}_{-31} \;MeV$ is by $16 \;MeV$ larger than
$\Lambda^{(3)}$.

The freezing value of $\alpha_B(0) = 0.5^{+0.06}_{-0.05}$ can be
found from the definition  Eq.~(\ref{eq29}) with $m_B=1.1\;GeV$.
The behaviour of $\alpha_B(\mu)$ as a function of $\mu$ is shown
in Figure~1. by the dashed curve. From Figure~1 one can see that
all seven values of $\alpha_s(\mu_0)$ found in our fit with
different $m$ lie on this curve $\alpha_B(\mu)$ in the region
$\mu \la 1.2 \;GeV$.

At the scale of the $\tau$ lepton mass the value of
$\alpha_B(M_{\tau}) = 0.306^{+0.015}_{-0.014}$ is close to the lower
limit of $\alpha_s(M_{\tau}) = 0.35 \pm 0.03$ from the compilation
\cite{1}. However, in some theoretical approaches, e.g. in
renormalon chain model, the value of $\alpha_s(M_{\tau}) = 0.305$
is preferred \cite{2}, which coincides with our prediction.

Note also that $\alpha_s$ freezing phenomena is actually
observed in recent lattice calculations of three gluon vertex function 
in \cite{25} where $\alpha_s(\mu)$ points at $\mu < 1.8 \;GeV$ 
lie far below than perturbative $\alpha_s(\mu)$ values and all 
points look like going to some constant value. 

\section{Conclusion}

Two fits -- to mass level differences and the fine
structure splittings in charmonium were done here. In all
calculations the relativistic kinematics was taken into account
which is more important for spin--dependent effects than for
the spin--averaged $c\overline{c}$ spectrum.

The fine structure parameters experimentally known from
precise measurements of $\chi_c$ masses are not small
$(\sim 35 \;MeV)$ and measured with accuracy better 2\%. The
value of linear combination $\eta = 3/2 c - a$ is also not small
and known with 5\% accuracy, the use of this combination allows
to simplify an extraction of $\alpha_s(\mu)$ from the fine
structure data.

For the fixed charm--quark mass $m$ and NP static interaction the
value of $\alpha_s(\mu_0)$ and the scale mass $\mu_0$ are
unambiguously calculated from our fits. In  practice the
variation of string tension is admitted in the range $0.17 \le
\sigma \le 0.185 \;GeV^2$ till a good fit to mass level
differences is obtained. The uncertainty in $\sigma$ value gives
rise to the theoretical error $\sim 5\%$ in $\alpha_s(\mu)$.

The dominant theoretical uncertainty comes from the choice of $c$
quark pole mass which has broader range than for $\overline{MS}$
mass, $\overline{m}(\overline{m}^2)$. Therefore we had to solve
spinless Salpeter equation for many sets of parameters with
different $m$.

The value of $m=1.48\pm0.08\;GeV$ was obtained in our best fit
which corresponds to the $\overline{MS}$ mass
$\overline{m}=1.18\pm0.07\;GeV$ and the extracted value of the
strong coupling constant is $\alpha_s(\mu_0)=0.38 \pm 0.03(exp)
\pm 0.04(theor)$ at scale $\mu_0 = 1.0 \pm 0.2\;GeV$ The
uncertainty connected with NP effects is included in theoretical
error. This value of $\alpha_s(1\;GeV)$ is rather close to that
calculated in bottomonium for $2P$ state \cite{4}, probably,
because $2P$ $b \overline{b}$ state has the same size as $1P$
state in charmonium.

We suggest to consider relatively small value of
$\alpha_s(1\;GeV)$ as a manifestation of $\alpha_s$ freezing and
compare our calculated values of $\alpha_s(\mu)$ at $\mu \sim
0.7 \div 1.2 \;GeV$ with theoretical formula of $\alpha_B(\mu)$
in background perturbation theory.

The behaviour of $\alpha_B(q^2)$ at large energy scale coincides
with perturbative predictions since $\Lambda_B^{(5)} =
\Lambda^{(5)}$ that gives rise to $\alpha_s(M_Z) = 0.119 \pm
0.002$ in two--loop approximation. For $n_f=3$ $\Lambda_B^{(3)} =
400^{+34}_{-32}\;MeV$ is slightly larger that perturbative
$\Lambda^{(3)} = 384.4 \;MeV$. With this $\Lambda_B^{(3)}$ the
predicted value  of $\alpha_s(M_{\tau}) \approx
0.306^{+0.015}_{-0.014}$ at the scale of $\tau$ lepton mass is
smaller that the conventional value $0.35\pm0.03$ but coincides
with the prediction in renormalon chain model.

The authors are grateful to Prof. B.~Bakker for possibility to
use his program code and Prof. Yu.A.Simonov for fruitful discussions.
This   work was partially supported by Russian Foundation for
Basic research (project N96-02-19184a) and the joint RFBR-DFG
foundation (grant N96-02-00088g).


\begin{figure}
\caption{ The running coupling constant as a function of the 
scale parameter $\mu$. The perturbative two--loop $\alpha_s(\mu)$ 
with 
$\Lambda^{(5)} = 237^{+26}_{-24} \;MeV$, 
$\Lambda^{(4)} = 338^{+33}_{-31} \;MeV$ and
$\Lambda^{(3)} = 384.4^{+32}_{-30} \;MeV$ is shown by a hatched 
band. The running two--loop coupling constant in background 
perturbation theory with the background mass $m_B = 1.1 \;GeV$, 
$\Lambda_B^{(5)} = \Lambda^{(5)}$,
$\Lambda^{(4)} = 339.2^{+33}_{-31} \;MeV$ and
$\Lambda^{(3)} = 399.6^{+34}_{-32} \;MeV$ is shown by a dashed 
lines. The values of $\alpha_s(\mu_0)$ extracted from the fit to
charmonium fine structure data for different quark masses are
depicted by open circles with error bars including both
theoretical and experimental uncertanties. }
\label{fig1}
\end{figure}

\newpage

\mediumtext
\begin{table}
\caption{The comparison of $1P$ state matrix elements for spinless
Salpeter and Schr\"odinger Equations.}
\begin{tabular}{|c|c|c|c|c|c|c|}
&\multicolumn{2}{|c|}{Set A} & \multicolumn{2}{|c|}{Set B}&
\multicolumn{2}{|c|}{Set C}
\\
matrix
&\multicolumn{2}{|c|}{\hspace*{0.8cm}$m=1.20\;GeV$\hspace*{0.8cm}}
&\multicolumn{2}{|c|}{\hspace*{0.8cm}$m=1.40\;GeV$\hspace*{0.8cm}}
&\multicolumn{2}{|c|}{\hspace*{0.8cm}$m=1.48\;GeV$\hspace*{0.8cm}}
\\
element &\multicolumn{2}{|c|}{$\sigma=0.20\;GeV^2$}
        &\multicolumn{2}{|c|}{$\sigma=0.17\;GeV^2$}
        &\multicolumn{2}{|c|}{$\sigma=0.18\;GeV^2$}
\\
        &\multicolumn{2}{|c|}{$\tilde{\alpha}_V=0.350$}
        &\multicolumn{2}{|c|}{$\tilde{\alpha}_V=0.373$}
        &\multicolumn{2}{|c|}{$\tilde{\alpha}_V=0.420$}
\\ \cline{2-7}
\rule{0pt}{15pt}&~R$~^{a)}$~&~NR$~^{a)}$~&~R~&~NR~&~R~&~NR~ \\ \tableline
\rule{0pt}{15pt}$\langle r^{-3} \rangle_{1P}(GeV^3)$
& \quad 0.117 \quad & \quad 0.080 \quad & \quad 0.118 \quad &
\quad 0.086 \quad & \quad 0.1423 \quad & \quad 0.1014 \quad \\ \tableline
\rule{0pt}{15pt}$\langle r^{-1} \rangle_{1P}(GeV)$
& 0.382 & 0.345 & 0.382 & 0.351 & 0.4040 & 0.3704 \\ \tableline
\rule{0pt}{15pt}$a_{NP}(MeV)$
&-26.53 &-23.96 &-16.57 &-15.22 &-16.60  &-15.22  \\ \tableline
\rule{0pt}{15pt}$\sqrt{\langle r^{2} \rangle_{1P}}(GeV^{-1})$
& 3.281 & 3.565 & 3.293 & 3.514 & 3.130  & 3.340  \\
\end{tabular}
\label{tab1}
a) R (NR) refers correspondingly to relativistic (nonrerativistic)
case.
\end{table}

\mediumtext
\begin{table}
\caption{The spin--orbit and tensor $1P$ state splittings and 
spin--averaged mass level differences in charmonium$~^{a)}$ (in $MeV$).}
\begin{tabular}{|c|c|c|c|}
& & Set C & Set D \\
& & $m=1.48\;GeV$, & $m=1.40\;GeV,$ \\
& Experimental & $\sigma=0.18\;GeV^2,$& $\sigma=0.183\;GeV^2,$\\
& values$~^{b)}$ & $\alpha_s(\mu_0)=0.365,$& $\alpha_s(\mu_0)=0.312,$\\
& $(MeV)$ & $\mu_0=0.909\;GeV,$& $ \mu_0=1.357\;GeV,$\\
& & $\tilde{\alpha}_V=0.42$
& $ \tilde{\alpha}_V=0.39$ \\
\tableline
\rule{0pt}{15pt}$a_{NP}$& --& $-16.60$& $-18.40$ \\
\rule{0pt}{15pt}$a_{P}^{(1)}$& --& $47.56$& $41.84$ \\
\rule{0pt}{15pt}$a_{P}^{(2)}$& --& $3.61$& $ 11.12$ \\
\rule{0pt}{15pt}$a_{total}$& $34.56\pm0.19$& $34.56$& $34.56$ \\
\tableline
\rule{0pt}{15pt}$c_{P}^{(1)}$& --&$31.70$& $27.89$ \\
\rule{0pt}{15pt}$c_{P}^{(2)}$& --&$7.42$& $11.23$ \\
\rule{0pt}{15pt}$c_{total}$& $39.12\pm0.62$& $39.12$& $39.12$ \\
\tableline
\rule{0pt}{15pt}$M(2S)-M(1S)$& $595.39\pm1.91$& $591.30$& $586.09$ \\
\rule{0pt}{15pt}$M(1P)-M(1S)$& $457.92\pm1.0$& $460.68$& $445.93$ \\
\rule{0pt}{15pt}$M(2S)-M(1P)$& $137.47\pm1.77$& $130.62$& $140.15$ \\
\end{tabular}
\label{tab2}
\end{table}
a) all matrix elements defining spin effects were calculated
   for spinless Salpeter equation.

b) $M(nL)$ means the spin--averaged mass.


\begin{thebibliography}{99}

\bibitem{1} 
Data compilations: Particle Data Group, The European Physical
Journal {\bf C~3}, (1998).

\bibitem{2} 
K.~Ackerstaff et.al., preprint CERN-EP/98-102
(submitted to Eur. Phys. Jour. C).

\bibitem{3} 
A.M.~Badalian and Y.P.~Yurov, Phys.~Rev. {\bf D~42}, 3138
(1990).

\bibitem{4} 
S.~Titard and F.J.~Yndurain, Phys.~Lett. {\bf B~351}, 541
(1995).

\bibitem{5} 
L.P.~Fulcher, Phys.~Rev. {\bf D~44}, 2079 (1991).

\bibitem{6} 
F.~Halzen, C.~Olson, M.G.~Olsson, and M.L.~Stong, Phys.~Rev. {\bf
D~47}, 3013 (1993).

\bibitem{7} 
S.~Titard and F.J.~Yndurain, Phys.~Rev. {\bf D~49}, 6007 (1994);
{\bf 51}, 6348 (1995). 

\bibitem{8} 
J.~Pantaleone, S--H.H~Tye, and Y.J.~Ng, Phys.~Rev. {\bf D~33}, 777
(1986).

\bibitem{9} 
S.~Jacobs, M.G.~Olsson, and C.~Suchyta, Phys.~Rev. {\bf D~33}, 3338
(1986).

\bibitem{10}
W.~Lucha, F.F.~Schoeberl, and D.~Gromes, Phys.~Rep. {\bf 200}, 127
(1991) and references therein.

\bibitem{11}
A.Yu.~Dubin, A.B.~Kaidalov, and Yu.A.~Simonov, Phys.~Lett. {\bf B~323},
41 (1994).

\bibitem{12}
M.~Peter, Phys.~Rev.~Lett. {\bf 78}, 602 (1997).

\bibitem{13}
H.G.~Dosch and Yu.A.~Simonov, Phys.~Lett. {\bf B~205}, 339 (1998); \\
A.M.~Badalian and Yu.A.~Simonov, Yad.~Fiz. {\bf 59}, 
2247 (1996); Phys.~Atom.~Nucl. {\bf 59}, 2164 (1996);\\
A.M.~Badalian and V.P.~Yurov, Yad.~Fiz. {\bf 56}, 239 (1993).

\bibitem{14}
C.~Michael et.al., Phys.~Lett. {\bf B~283}, 103 (1992); \\
S.P.~Booth et.al., Phys.~Lett. {\bf B~294}, 385 (1992); \\
G.S.~Bali and K.~Schilling, Phys.~Rev. {\bf D~47}, 661 (1993).

\bibitem{15}
A.~Di Giacomo, M.D.~Elia, H.~Panagopoulos, and E.~Meggiolaro, 
hep-lat/9808056.

\newpage

\bibitem{16}
F.J.~Yndurain, hep-ph/9708448; \\
R.~Akhoury and V.I.~Zakharov, Phys.~Lett. {\bf B~438}, 165 
(1998).

\bibitem{17} 
YU.A.~Simonov, in preparation.

\bibitem{18} 
A.M.~Badalian and V.L.~Morgunov, Yad.~Fiz. (in press).

\bibitem{19} 
K.D.~Born et.al., Phys.~Lett. {\bf B~329}, 332 (1994); \\
G.S.~Bali, K.~Schilling, and A.~Wachter, Phys.~Rev. {\bf D~56}, 
2566 (1997).

\bibitem{20}
Yu.A.~Simonov, Pis'ma~Zh.~Eksp.~Teor.~Fiz. {\bf 57}, 513 (1993); \\
Yad.~Fiz. {\bf 58}, 113 (1995).

\bibitem{21} 
A.C.~Mattingly and P.M.~Stevenson, Phys.~Rev. {\bf D~49}, 437 (1994).

\bibitem{22} 
A.M.~Badalian and Yu.A.~Simonov, Yad.~Fiz. {\bf 60}, 714 (1997).

\bibitem{23} 
A.M.~Badalian, Phys.~Atom.~Nucl. {\bf 60}, 1003 (1997) 
(translated from Yad.~Fiz. {\bf 60}, 1122 (1997)); 
hep-lat/9704004.

\bibitem{24} 
M.~Luescher, Nucl.~Phys. {\bf B~180}, 317 (1981);  \\
S.~Perantonis and C.Michael, {\bf B~347}, 854 (1990); \\
T.~Manke et al. CP-PACS Collaboration, hep-lat/9812017.

\bibitem{25} 
G.~Burgio, F.~Di~Renzo, C.~Parrinello, and C.~Pittori, 
hep-ph/9809450, hep-ph/9808258.


\end{thebibliography}
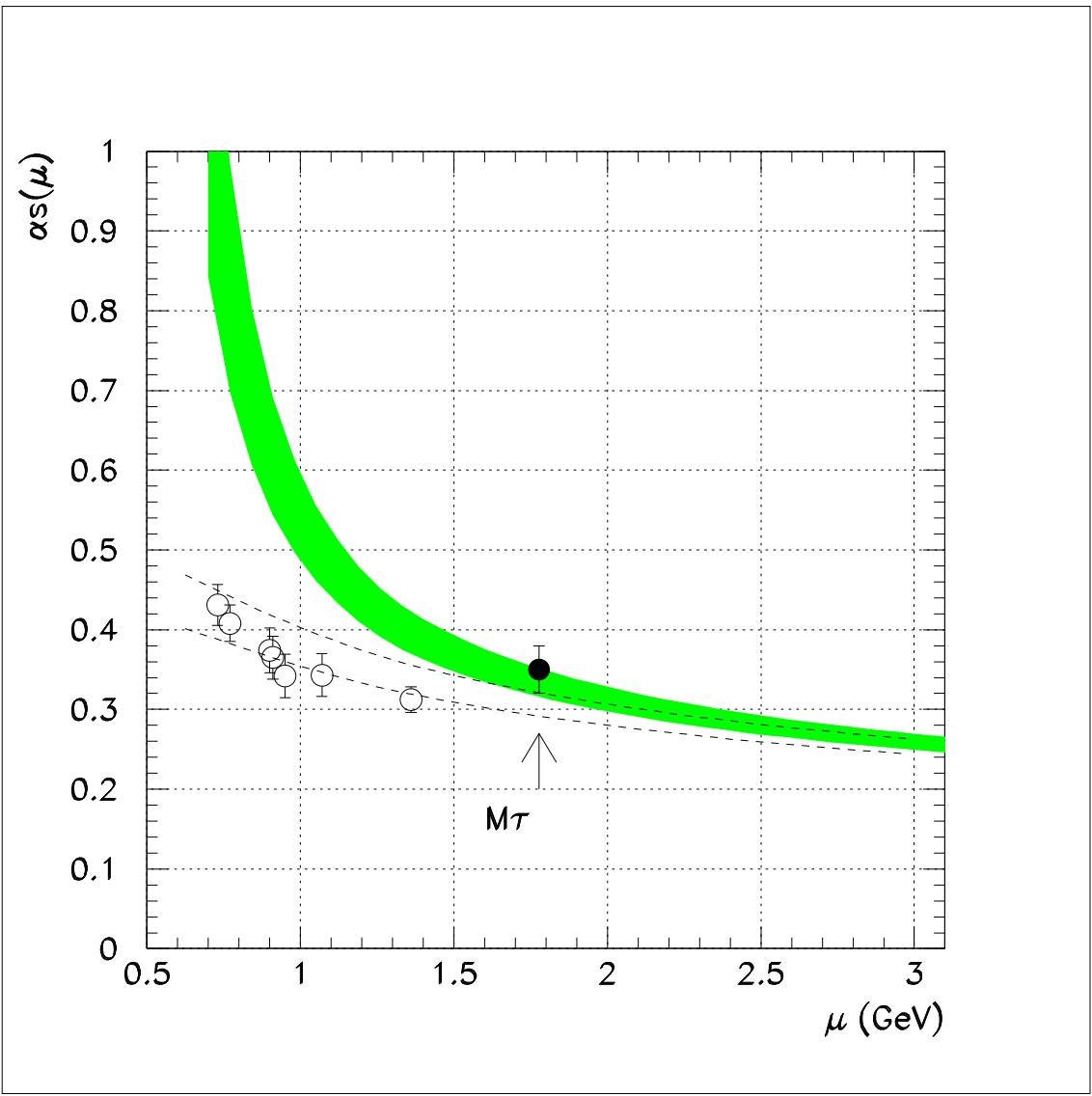
\end{document}